# Distributed archive and single access system for accelerometric event data: a NERIES initiative

**Catherine Péquegnat**[1], **Raphael Jacquot**[1], **Philippe Guéguen**[1], **Stéphanie Godey**[2], **and Laurent Frobert**[2]

[1] LGIT/CNRS/LCPC - Joseph Fourier University Grenoble, France.

[2] EMSC/CSEM -European-Mediterranean Seismological Centre, Bruyères-le Châtel, France.

**Abstract.** We developed a common access facility to homogeneously formatted accelerometric event data and to the corresponding sheet of ground motion parameters. This paper is focused on the description of the technical development of the accelerometric data server and the link with the accelerometric data explorer. The server is the third node of the 3-tier architecture of the distributed archive system for accelerometric data. The server is the link between the data users and the accelerometric data portlet. The server follows three main steps: (1) Reading and analysis of the end-user request; (2) Processing and converting data; and (3) Archiving and updating the accelerometric data explorer. This paper presents the description of the data server and the data explorer for accessing data.

## Introduction

One activity of the NERIES project is to improve the access to the European accelerometric data. The necessity is the development of common access to homogeneously formatted accelerometric data and to the corresponding sheet of ground motion parameters. First, the input and output formats, the protocols for exchanging data, the structures of the database and the description of acquisition channels were discussed and agreed on with the other partners. A preliminary version of the accelerometric database structure was defined based on the experience of the French Accelerometric Network database [1]. Structures and formats of the database tables were fixed in relation to the description of soil conditions and EMSC inventory of the accelerometric stations in Europe [2]. Based on the inventory, the description of the accelerometric channels found in Europe were implemented in conformity with the tool used for building SEED volume.

The distributed system for accelerometric event data was described in 2007 in Gueguen et al. [3], the prototype data server implemented in 2009 and the accelerometric data explorer as a portlet in the seismic data portal in 2009. The data server (invisible for end users as well as for data providers) is an important piece of the system: in charge of collecting, processing and archiving the data, it will allow their dissemination through a unique and standard system. Unlike the broad-band community which has defined standards and protocols for exchanging data a long time ago, accelerometric networks are less structured and the data server was a good opportunity to define and fix some specifications.

Based on the French Accelerometric Network experiences, conversion tools from earthquake engineering standardized format (ASCII) to seismology standardized format (SAC, miniseed) was chosen in order to increase the dissemination of these data to a large community. For this reason, the accelerometric system has been defined following a 3-tiers architecture: data providers (accelerometric networks) in charge of giving access to their data and metadata; data explorer to the data for the management of the end-users requests; and the data server only focused on the processing, the conversion and the download of the data, following the request file transmitted by the accelerometric data explorer.

The accelerometric data explorer and the data server are described in this document. The main objectives of the data server are first described, as well as the structure chosen for its development. During the entire period of the project, the main concern was to select and choose solutions as close as those existing for the other system devoted to the data access of the seismological data. The metadata stream between accelerometric data explorer and data server is then described in the second part of this document. Finally, the specific tools developed are listed and described, from the XML files reader to the data converter.

## Distributed system

As described in Guéguen et al. [3], the specifications of the system for accessing the data have been defined in three parts:

> 1. Data are waveforms (accelerograms) and engineering parametric data. Moreover, the data are "event based", that means that they are explicitly linked to events by data providers.
>
> 2. Data selectors and data extraction tools will allow the expression of two kinds of criteria for retrieval: seismological criteria (event, magnitude, station to event distance) and engineering criteria (PGA, PGV etc).
>
> 3. Accelerometric data are delivered by tools in 2 formats :
>
>> - in a specific ASCII format for the engineering community,
>> - in SEED format in order to share and disseminate it widely.

### Specific ASCII format for accelerometric data

Different formats have been examined from Strong Motion (SM) databases: COSMOS (The Consortium of Organizations for Strong-Motion Observation Systems, www.cosmos-eq.org) in California, KNET in Japan [4], and the RAP in France [1]. All of them are composed by i) headers containing information related to event, station and record and ii) acceleration time histories.

Some basic ideas have guided our choices:

> 4. a simple header with the most important information concerning station, event and record. A similar format to the one used by RAP seems to be the most adequate, between the very simple content of KNET and the comprehensive content of COSMOS (result of a long history of US formats).
>
> 5. waveforms will be present by one sample per line with a first column for time and a second for acceleration
>
> 6. SM parameters and PSV response spectra (5% damping) at 28 frequencies computed and stocked by each participant in a homogeneous way [5] and available in the data explorer are also provided.

In conclusion, data in ASCII format will be split into 2 volumes: the first one for accelerations and the second one for parameters (SM parameters and PSV (5%) response spectra at 28 frequencies) with the same header. A common header for metadata will be added to each volume.

**Main nodes**

The distributed system is based on a 3-tier architecture, the nodes of which are (1) data providers' nodes, (2) data explorer and metadata node and (3) data server node (Fig. 1).

> 7. Data providers' nodes do set up ftp server offering protected access to primary waveforms (without any header) and partial metadata in two ASCII tables. The first table is for event-record association, the second is for record parameters. The syntax and the semantic of these two ASCII tables: *event-record.txt* and *parameters.txt,* as well and the directory structure for the initial raw data files have been defined in Guéguen et al. [3].
>
> 8. EMSC node provides User Interfaces for data queries expression. EMSC operates two databases: the EMSC event database and the station metadata database, which contains all the metadata about networks, i.e. site conditions, stations, channels, records and parameters, except the specific information about instrumental responses. The structure of the station metadata database (which is derived from the RAP database) has been described in Guéguen et al. [3]. The content of this database is based on the EMSC report [6].
>
> 9. The last node is the data server, which processes queries and data. The main ressources of this node are :
>
>> • A local metadata database, the structure of which is very close to the metadata database. This local database will be progressively filled up with and during requests processing.
>> • The PZ database (poles and zeroes database) which contain the generic instrumental responses of the NA5 instruments (8 types of digitizer configurations, 7 types of generic sensor). The structure of the PZ database, its content as well as the tools used to manage it have been described in details in Guéguen et al. [3].

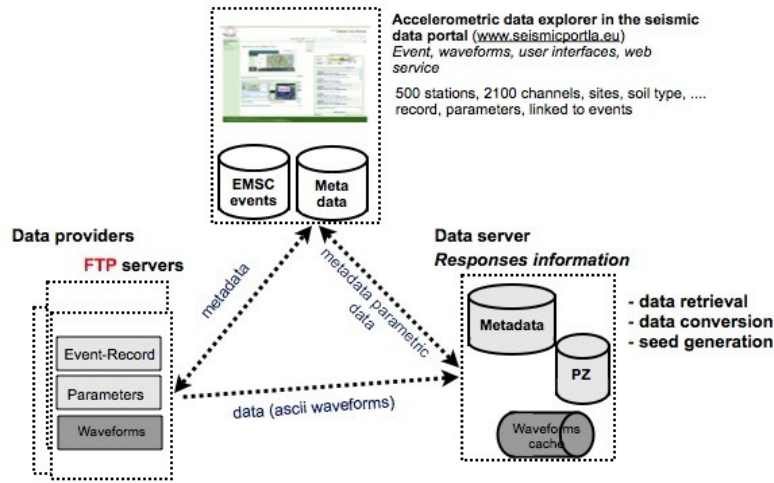

**Fig. 1.** Description of the three nodes of the system, between the accelerometric data explorer, the data providers and the data server.

### Metadata stream between the accelerometric data explorer, data provider and data server

The metadata stream between the accelerometric data explorer and the data providers' nodes is the following (Fig. 2):
- Providers should use a Webservice to retrieve hypocentral localisations (in XML format) from the EMSC database. In this database and in the retrieved XML file, events are identified by a key called UNID. The UNID implementation by the EMSC is described in [2] and [6].
- Providers have then to link their waveforms to this UNID using their own association methods and store the result in the formatted ASCII file: *event-record.txt*

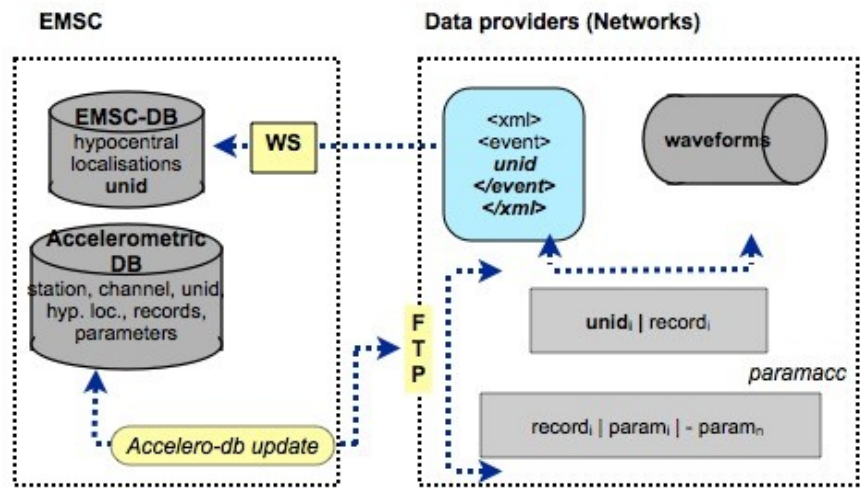

**Fig. 2.** Description of the stream between the accelerometric data explorer in the seismic data portal (www.seismicportal.eu)

The *event-record.txt* ASCII table is the input of the PARAMACC software [5], which computes engineering parameters for each record, and stores the values in a second ASCII table: *parameters.txt*. Those two tables must be accessible on the ftp server of the providers, known by the data explorer. It downloads ASCII tables and integrates them into the metadata database.

The metadata stream between data explorer and data server is the following (Fig. 3):
- User queries, built on the content of the metadata database, are translated to an *XML* file, which will be pushed to the data server using a messaging queue system (RabbitMQ).
- Once on the data server, the XML file is loaded into an internal database before being treated by the service in order to build the archive for the user.
- At the end of the process, the data server will rewrite the XML file (filling some more fields stored in the local database during the request processing such as for instance the exact size of the record) and push it back to the data explorer node.
- It is the task of the data explorer to notify the user that the archive is ready to pick up on the ftp site of the data server.

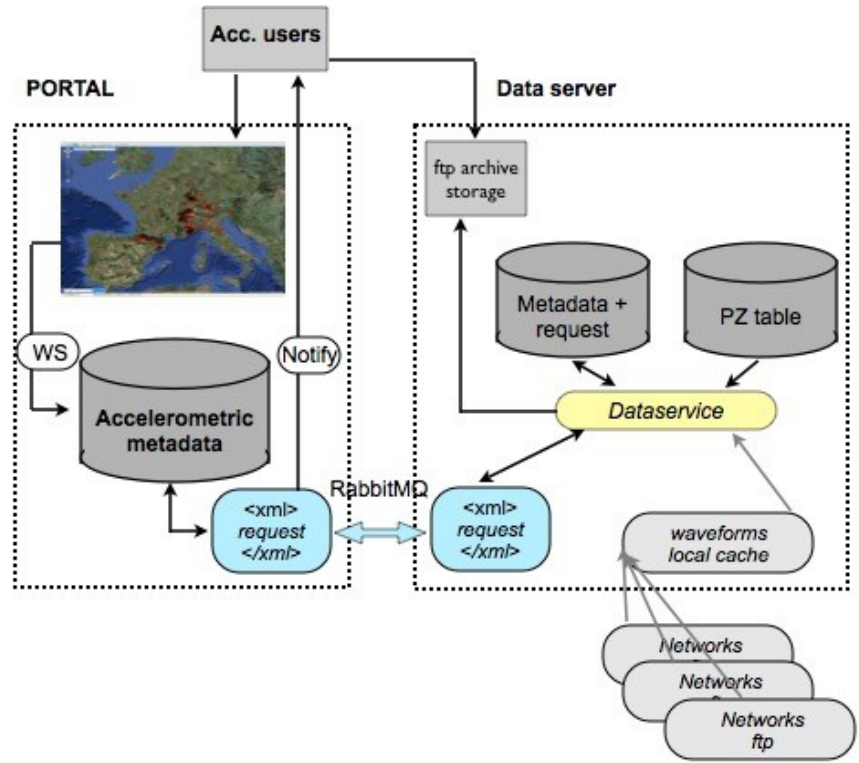

**Fig. 3.** Description of the stream between the data explorer and the accelerometric data server

## Detailed Data server description

### Structure of the request

The accelerometric data explorer sends requests to the data server in the form of an XML file that contains the required data in a structured form. All the information concerning the XML file structure is described in Péquegnat et al. [7].
This XML file contains a series of sections, including information about the request itself, the various organizations involved in the stations related to the request, the various networks to which the stations and sensors belong, a description of all hardware types, the list of events that have been identified as pertaining to the request, the description of the stations, and the complete descriptions of all records that have been found as corresponding to the request. Some flag are added to the XML file of the response, some field found in the XML request files being preserved in the response without modification, some added to the XML response files calculated or filled in by the data server.

*Request section***:** This section is the root of the XML document, including information about the requested data and the requesting user and filled in by the accelerometric data explorer: it contains a unique identifier for the request, determined by the data explorer, the types of requested data (waveforms, parameters or both), the format of the output data (ASCII, SAC, SEED, miniSEED), the email address of the user and some information on the success or not of the request.

*Organization section*: This section describes either a research facility, or a hardware / software manufacturer. The attribute of the section are a unique ID for the organization and the URL of the organization website.

*Network section***:** this section describes the networks, owner of the data, found in the request. The attribute is a unique ID for the network, the FDSN unique ID for the network (2-letter codes used in seed data)**,** a time stamp that describes when the network was started, an optional time stamp that describes when the network was stopped, the URL to the network's website (which should contain a description of the network)**,** the organization ID pointing to the organization that manages the network. The FDSN code attribute is optional, because all the accelerometric networks do not have such a code. In that case, a default code will be added by the data server tool in order to deliver seed volume anyway.

*Equipment section***:** This section has no attributes and contains one or more "sensor", "preamp" and "adc" elements. The "***sensor***" element describes each type of sensor found in the request. It contains a unique ID for the sensor type, given by the data explorer, the ID of the manufacturer that refers to the ID of an "organization" element, the URL to the file with the PZ description and values for the sensor (pz_url), the corner frequency of the sensor, the seed code for the sensor band and the sensor sensitivity. In the incoming XML file, the pz_url attribute is empty. It will be filled up by the data server tools. The rule to derive the pz_url base name file is to concatenate the manufacturer and the sensor ID. If the pz_url element remains empty in an outgoing XML file, it signals an inconsistency in the PZ database. The "***preamp***" element describes each type of preamplifier found in the request, in relation with the sensor. It contains a unique ID for the preamplifier type, generated by the data explorer application, the URL to the file with the PZ values for the preamplifier**,** the value of the analogue gain for the preamplifier. The "***adc***" element describes each type of digitizer found in the request. It is formed by an unique ID for the pair type of digitizer code – sampling frequency, generated by the data explorer application, the ID of the manufacturer (refers to the ID of an "organization" element)**,** the digitizer type code, the URL to the file with the PZ values for the digitizer**,** the sampling frequency, in Hz, and the digital gain as set on this digitizer type. In the station metadata database, ***adc id*** are integer (numeric keys) and a foreign key give access to the data type code via the dastype table. ADC entries have 3 dependent fields: the analogue device converter code (ADC), the sampling frequency and the numerical gain. In case of an unknown ADC, the 'XXXX' word will be used.

*Event section*: this section describes the events that are requested. It contains the UNID of the event provided by EMSC/CSEM, the date of event, the longitude, latitude and depth of the epicenter and the standardized name of the region where the event occurred.

Complementary to the event section, the *magnitude section* is defined for each event element. It contains the type of the magnitude value and the value of the magnitude of the event, each event may have several magnitude values ($M_L$, $M_w$, mb etc...).

*Station section*: this section describes the stations found in the request. It contains the station ID used in seed data, the URL at which the station is described, the station's owner ID (referring to one of the "organization" elements above), the station's manager ID (referring to one of the "organization" elements above), the soil type where the station is located ("S" or "R"), the Eurocode definition for the soil where the station is located, the latitude and longitude of the station.

*Channel section*: this section describes all the information linked with the data. For each channel, there's a "site" description, an "equipment" description and a list of "record". The attributes are the channel code used in seed data (HNN for example), the channel's location code (01 for example), the channel's azimuth, a time stamp to identify the day when the channel was set up, a time stamp to indicate the day when the channel was stopped, a distance in meters where to the north the channel sensor is located with respect to the station, a distance in meters where to the east the channel sensor is located with respect to the station, a distance in meters of how much higher the channel sensor is located with respect to the station (ground level), the depth at which the sensor is located with respect to the above mentioned ground level, the URL of the web page where the channel is described, an optional URL to the channel's response file, an optional URL to the channel's dataless file (containing the metadata), the channel's manager (refers to an organization ID) and the channel's seed network ID.

*Site section*: this section describes the building where the channel is located. The attributes are the type of the site, the total number of floors in the building, the floor on which the sensor is located and the depth at which the sensor is located with respect to the floor. The site types are:

- R: rock based on surface geological observations
- A: rock or stiff geological formation (Vs30>800 m/s)
- B: stiff deposits of sand, gravel or over consolidated clays (Vs30>360m/s and Vs30<800 m/s)
- C: deep deposit of medium dense sand, gravel or medium stiff clays (Vs30>180m/s and Vs30<360 m/s)
- D: loose cohesionless soil deposits (Vs30<180 m/s)

- E: soil made up of superficial alluvial layer, with a thickness ranging from 5 to 20m with Vs30 value in class C and D ranges covering stiffer deposits (class A)
- BHd: Borehole sensors. d=depth in meters
- Bm.n: Sensors in building (m=total number of stories of the building including ground floor, n=floor where the instrument is installed)
- O: other

*Record section*: this section contains the information on the record in relation with the channel and the event. The attributes are the event ID (refers to the UNID of one of the events that is requested), the URL of the source data for this record, the URL for the corresponding miniSEED file, the size of the data in bytes, the time stamp of the record, the time stamp of the first sample in the record that corresponds to the event, the time stamp of the last sample in the record, the time stamp of when the record was created, the PGA_uncorrected parameter, the PGA_corrected parameter, the Arias intensity, the Trifunac parameter, the CAV parameter, the PGV parameter, the 28 values defining the PSV parameter and the Housner parameter.

**Local database**

A database is used locally to store the information from the requests and responses. This database contains all the fields defined in the XML files described above. When a processing task has finished, it updates the database to reflect the status of the data.

At the end of the request, the last process extracts the information from the database to create the response XML file that is sent back to the data explorer.

**Application Architecture**

The application running on the data server, that is responsible for generating the request response files containing the requested data, is designed as a pair of processes that run in parallel as system daemons (Fig. 4). A request-reader task watches over a directory, waiting for request description XML files to come in, then reads it into the database, and signals via RabbitMQ the second module, the data-engine, to do the actual work. Each module is a thread (that is, it runs in parallel to other modules if the data it requires is ready at the time). The data-engine starts the various modules, which start doing their work and signal each other to synchronize the work to be done. All the information concerning the application architecture are described in Péquegnat et al. [7].

**Request-reader.py**

The request-reader.py application is used at two locations in the system. It is used as the loader application. When started with the proper arguments, this application uses the notify system call to watch a directory in which RabbitMQ is configured to drop the XML request files when they arrive. The application then opens the XML file, and uses the libXML reader interface to parse the XML file into its components, and create specific python objects for each XML tag. The parsing follows instructions contained in a python dictionnary data structure (NA5RE-QUEST). Those instructions include information about the fact that the parameter is required, the parameter type and list of valid values. It also contains pointers to a pair of functions (get and put) that know how to persist the object in the local database.

One of the most important functions in the application is the parse function. This function does the final action of the work, including creating each object. This function is also called recursively by the objects themselves when time comes to parse the XML tags included as children of those objects.

The read_input function opens the file, creates the libXML reader interface to it, and does the first call to the parse function. If the appropriate option was given, this function also sends the dbus message alerting that the request is completely loaded in the database and ready to be processed.

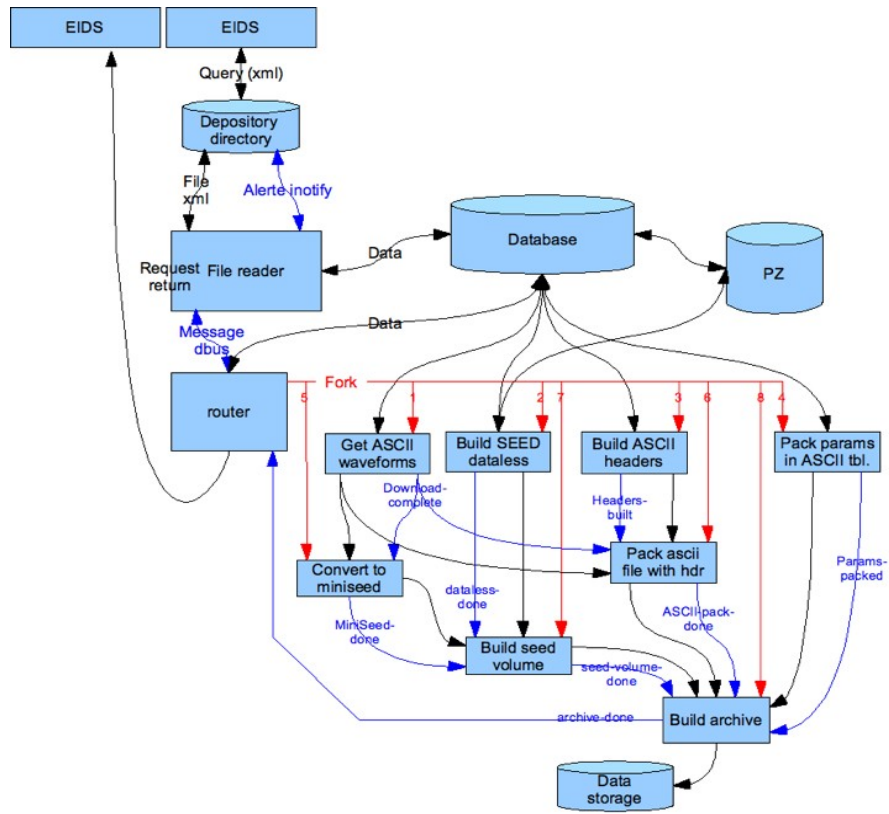

**Fig. 4.** Description of the accelerometric data server designed for processing the request coming from the accelerometric data explorer.

The write_request function gets the request from the database and outputs it as an XML file. This function is used at the end of the process to send the data to the data explorer. The watch_directory function uses the notify system call to wait for an XML file to appear in from EIDS. Once the file is located, this calls the read_input function with the proper parameters. The main program handles parameter analysis, sets the appropriate options, and calls one of the above functions depending on the command line options that were passed.

*data-engine.py:* This process runs in the background, waiting for a DBus message from the request-reader task. Once the message is received, the data-engine starts a Laucher thread that will in turn start all the other processes required to execute the request.

*downloader.py:* This thread downloads the original ASCII files that are not cached yet locally. If the download is successful, the local URL is stored in the database (which signals the presence of the file for subsequent runs. When all files are downloaded, the thread signals that it has finished, which in turn allows for subsequent tasks to start. If a file can't be downloaded, then the URL field in the database is empty, warning the failure.

*datalessmaker.py:* This task generates a SEED dataless file from the data sent in the request for all the channels of each station it contains. It widely uses some tools coming from the BDsis database system set up for the RAP network [1]. The way the datalessmaker works is the following:

1. An (ASCII) intermediary file is built for each station, using some pieces of the specific channels information stored in the internal database (and particularly but not only, the pz_url attributes of the *sensor*, *preamp* and *adc* elements). Such a file (called 'dbird', for 'database instrumental responses description') establishes all the links between the different channels elements for a station, and the corresponding (relative) PZ files of the PZ database. An example is given further.
2. The seed writer itself reads a dbird file, loads the entire PZ database, checks that the PZ files names which are referenced in the dbird file are available, performs some controls in those PZ files, and build the dataless volume. The seed writer first builds Stations header logical records, then builds Abbreviations header logical records, and finally builds Time header logical records.

*sacmaker.py:* Once the ASCII data is downloaded, this thread will transform each file into a file in the SAC format. It is run twice, the first time to generate ASCII format SAC files, the second time, to generate Binary format SAC files. This thread uses a pure python implementation of SAC writing that can be found in pysac.py.

*mseedmaker.py:* This task converts the binary SAC files built above into mini-SEED files using the sac2mseed program. The main part is identifying which SAC files are there that haven't been yet transformed into mini-seed, then calling sac2mseed for each of them. Once this is done, the task alerts whichever other task that is interested that it is finished and that they can start their own work.

*seedmaker.py:* This task build a large SEED file for each station with all the mini-SEED info previously set up. The miniSEED files are 'cat-ed', and the seed volume for each station is built using the RDSEED program.

This task generates the appropriate ASCII data files according to the 2 volumes

```
volume 1                                          volume 2
---- EVENT                                        ---- EVENT
unid : <unid>                                     unid : <unid>
source : EMSC/CSEM                                source : EMSC/CSEM
localisation : NW SPAIN                           localisation : NW SPAIN
1996-02-18 01:45:45.5 2.54 42.79 8 5.2            1996-02-18 01:45:45.5 2.54 42.79 8 5.2
  <type_mag1> 6.3 <type_mag2> <mag3> <type_mag3>    <type_mag1> 6.3
---- STATION                                      ---- STATION
code : <stationcode> network : IGC                code : <stationcode> network : IGC
latitude : 42.730 longitude : 7.320 elevation : 250.0  latitude : 42.730 longitude : 7.320 elevation : 250.0
depth : 0.000 site : <R/S/A/B/C/D> sensor : EST   depth : 0.000 site : <R/S/A/B/C/D> sensor : EST
building_code : XX                                building_code : XX
---- EVENT-STATION                                ---- EVENT-STATION
distance_to_event : 68.0                          distance_to_event : 68.0
event_to_station_azimuth : 206.16                 event_to_station_azimuth : 206.16
station_to_event_azimuth : 25.87                  station_to_event_azimuth : 25.87
---- RECORD                                       ---- RECORD
julian_day : 327/2006 time : 06:58:01             julian_day : 327/2006
sample_rate : 0.00800 nb_points : 5632            component : ENE
component : ENE unit : cm/s2                      azimuth : 90.0 incident : 90.0
azimuth : 90.0 incident : 90.0                    ---- RECORD-PARAMETERS
                                                  pgau : 4.573729 cm/s2
-----------------------------------------         pga : 4.597282 cm/s2
VERSION : 2008-07-24T12:35:21Z – Neries project   al : 0.011488 cm/s
-----------------------------------------         td : 8.310000 s
0.0000 1.527225e-02                               cav : 5.622970 cm/s
0.0080 6.927599e-03                               pgv : 0.198258 cm/s
0.0160 -6.662256e-03                              hl : xxxxx cm
                                                  f(hz) psv(cm/s)
                                                  0.15 0.594359
                                                  0.19 0.297106
                                                  0.23 0.219786
                                                  0.28 0.185797
                                                  0.34 0.220771
                                                  0.42 0.298527
                                                  0.52 0.109993
                                                  [… 28 values total]
                                                  -----------------------------------------
                                                  VERSION : 2008-07-24T12:35:21Z – Neries project
                                                  -----------------------------------------
```

**Fig. 5**. Description of the Volume1 and Volume 2 of the ASCII format.

format as shown in Fig. 5.

## The accelerometric data explorer

The accelerometric data explorer is part of the seismic data explorer developed within the NERIES project and is accessible at www.seismicportal.eu. This portal is an aggregation of different portlets (thought as different applications) running on the server side from different Web servers but visually on the same Web Portal, accessible at an unique web address (Fig. 6).

This Portal currently includes:
- The Event Explorer to search for event information;
- The Wave Form Explorer for retrieving Broad band waveform data;
- Access to historical events' data;
- The accelerometric data explorer as presented here.

This last portlet is a visual front end and easy tool to search for accelerometric data using specific criteria (upon specific seismic events or accelerometric para-meters). This portlet helps the end-user to have access to accelerometric data using a single access point giving access to accelerometric data providers (currently six). The back end of the portlet consists of a database of the station metadata (network, hardware, sensors installation ...), a database of event records with accelerometric parameters and the events catalog from the data explorer.

The portlet is used to query these databases to found existing record event and retrieving accelerometric data in an easy user interface. An example of the user interface is given on Figure 7.

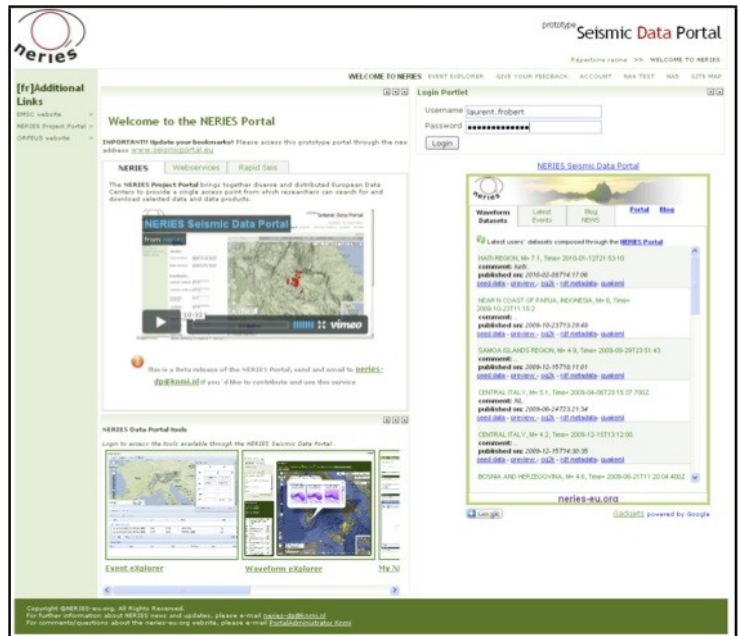

**Fig. 6.** The Neries Web portal access (**www.seismicportal.eu**). Access to the Accelerometric portlet is available for registered users.

**Fig. 7.** The Accelerometric Data Explorer portlet shows a map (1) displaying the result of the user query. A user query is made using the Accelero Search Criteria (2) The Waveform chooser (3) displays all the records available for the selected events, the user chooses the desired records, give a name to the query, selects the result format and sends the query. In the Queries panel (4) the user can see the different queries made and the current status of each query. When a query has been processed a link to the result is provided and the user downloads the result.

**Conclusion**

A first and preliminary version of the database used for the accelerometric data server and the accelerometric data explorer has been defined, as a portlet in the NERIES seismic data portal. The data server was designed for uploading data (waveforms and ground motion parameters) in several format (SAC, ASCII and MiniSEED). The tier architecture of the system designed for accelerometric data was defined for seperating the data providers, the data server and data explorer system. The stream between the data explorer and the data server is based on an XML formatted file containing all the informations.

The inventory and the desciption of the accelerometric stations in the Euro-Med region is used by the data explorer for helping the end-users of the accelerometric data to perform request. The data server and the data explorer collect, archive and provide data from several European accelerometric networks (data providers) in the homogeneous format and ground motion parameters. By this way, it is now possible to download accelerometric data corresponding to seismic events localized at the borders of several countries and recorded by several European networks.

Currently, data from the partners (IGC Barcelona, IST Portugal, RAP-LGIT Grenoble, ETH Zurich, ITSAK Greece and KOERI Turkey) are included in the data server, concerning data with magnitude over 2 since 2000. In the next step we plan to integrate data from other European networks as actually in progress for the Italian accelerometric data.

**Acknowledgement**. The work presented in this article is part of the EC Project NERIES, Sixth Framework Programme, Contract number: RII3-CT-2006-026130.

# References

[1] Péquegnat, C., Guéguen, P., Hatzfeld, D., Langlais, M. (2008) The French Accelerometric Network (RAP) and National Data Centre (RAP-NDC), *Seism. Res. Lett.*, 79(1), 79-89.
[2] Godey, S., Bossu, R., Guilbert, J. and Mazet-Roux, G. (2006) The Euro-Mediterranean Bulletin: A Comprehensive Seismological Bulletin at Regional Scale. *Seism. Res. Lett.*, 77(4), 460-474.
[3] Guéguen, P., Péquegnat, C. and Revilla, J. (2007) Specifications of protocols and formats for waveforms access. Task C - Delivrable D3 of the Networking Activity 5, NERIES project (EC project number 026130), Sixth Framework Programme EC, 38 pages.
[4] Okada, Y., Kasahara, K., Hori, S., Obara, K., Sekiguchi, S., Fujiwara, H. and Yamamoto, A. (2004) Recent progress of seismic observation networks in Japan (Hi-net, F-net, K-NET and KiK-net). *Earth Planets Space*, 56, xv–xxviii.
[5] Tapia, M., Susagna, T. and Goula, X. (2007) Task B - Specifications for PSA and PSV - Definition and computation of parametric data. Delivrable D2 of the Networking Activity 5, NERIES project (EC project number 026130), Sixth Framework Programme EC,140 pages.
[6] Godey, S. (2007) Define and maintain accelerometric station metadata , Delivrable D1 of the Networking Activity 5, NERIES project (EC project number 026130), Sixth Framework Programme EC, 36 pages.
[7] Péquegnat, C., Jacquot, R. and Guéguen, P. (2009) Development and implementation of unified waveform request. Task C - Delivrable D5 of the Networking Activity 5, NERIES project (EC project number 026130), Sixth Framework Programme EC, 91 pages.